\documentclass[12pt]{article}
\def\half{{\scriptstyle\frac{1}{2}}}
\begin{document}
\begin{center}
{\Large\bf Green's function for longitudinal shear of a periodic laminate}\\
\vskip .25in
{\sl John Willis\\
DAMTP, Cambridge}
\end{center}
\vskip .25in
The reasoning to follow is a slight extension of that employed in my article [1] and its
further development [2]. A formal solution to the title problem is presented for a general
periodic laminate and its application for the construction of ``effective constitutive relations'' is developed. Its implementation in detail will be reported separately.
\section{Exact Green's function}
The medium occupies all of space and its properties are defined by the periodic functions
$\mu^0_{13}(x_1)$, $\mu^0_{23}(x_1)$ and $\rho^0(x_1)$ which have period $h$. The medium
is considered to be random, in the sense that its shear moduli and density at position
${\bf x} = (x_1,x_2,x_3)$ take the values
\begin{equation}
\mu_{13}({\bf x},y) = \mu^0_{13}(x_1+y),\;\;\mu_{23}({\bf x},y) = \mu^0_{23}(x_1+y),\;\;
\rho({\bf x},y) = \rho^0(x_1+y),
\end{equation}
where the sample parameter $y$ is uniformly distributed on the interval $(0,h)$.

Now consider the application of a body force ${\bf f}({\bf x})$, with $i$-component
\begin{equation}
f_i({\bf x},t) = \delta_{i3}\delta(x_1)\delta(x_2)\delta(t).
\end{equation}
It generates just a $3$-component of displacement, $u_3(x_1,x_2,t,y)$, which satisfies
the equation of motion
\begin{equation}
\sigma_{13,1} + \sigma_{23,2} + \delta(x_1)\delta(x_2) = \dot p,
\label{motion}\end{equation}
where
\begin{equation}
\sigma_{13} = \mu_{13}w_{,1}\;\;\sigma_{23} = \mu_{23}w_{,2}\;\hbox{ and }\; p = \rho\dot u_3.
\end{equation}
Henceforth, equations will be considered in which Laplace transforms have been taken with
respect to $t$ and $x_2$, which is equivalent to considering dependence $e^{(st + k x_2)}$.
The complex variable $s$ is taken to have a small positive real part, to ensure causality.

Expressed in terms of $u_3$, the equation of motion (\ref{motion}) becomes
\begin{equation}
\{\mu^0_{13}(x_1+y)u_3^\prime(x_1,y)\}^\prime + k^2\mu^0_{23}(x_1+y)u_3(x_1,y) + \delta(x_1)
= \rho^0(x_1+y)s^2 u_3(x_1,y),
\label{motion2}\end{equation}
where the prime represents differentiation with respect to $x_1$. Equivalently, with
\begin{equation}
\overline u_3(x_1,y) = u_3(x_1-y,y)\;\hbox{ so that }\;u_3(x_1,y) = \overline u_3(x_1+y,y),
\label{uubar}\end{equation}
the equation of motion can be written
\begin{equation}
\{\mu^0_{13}(x_1)\overline u_3^\prime(x_1,y)\}^\prime + k^2\mu^0_{23}(x_1) \overline u_3(x_1,y)
+ \delta(x_1-y) = \rho^0(x_1) s^2\overline u_3(x_1,y).
\label{motion3}\end{equation}

The solution of equation (\ref{motion3}) can be expressed in terms of the two ``Floquet''
solutions,
\begin{equation}
\phi_+(x_1) = \psi_+(x_1)e^{\kappa x_1},\;\;\;\phi_-(x_1) = \psi_-(x_1)e^{-\kappa x_1},
\end{equation}
where $\psi_\pm$ are periodic with period $h$ and $\kappa$ has negative real part.\footnote{This
convention is a little different from that used in \cite{JRW09},
in which a parameter $\mu$ is used for
$\kappa$, and $\mu$ was defined to have positive imaginary part. There are corresponding minor
differences in subsequent formulae. The present convention
is what I used in a program written for oblique waves.}
Correspondingly, $\phi_+ \to 0$ as $x_1\to +\infty$ and $\phi_-\to 0$ as $x_1\to -\infty$.

Then,
\begin{equation}
\overline u_3(x_1,y) = \cases{D\phi_+(x_1)\phi_-(y),\;\;x_1 \geq y,\cr
                              D\phi_+(y)\phi_-(x_1),\;\;x_1\leq y,}
\end{equation}
where
\begin{equation}
D = \{\mu^0_{13}y)[\phi_+(y)\phi_-^\prime(y)-\phi_+^\prime(y)\phi_-(y)]\}^{-1}
\end{equation}
(which, in fact, is independent of $y$). The desired Green's function $G(x_1,y)$ is just
$u_3(x_1,y)$. Thus, it follows from (\ref{uubar}) that
\begin{equation}
G(x_1,y) = \cases{D\phi_+(x_1+y)\phi_-(y) \equiv D\psi_+(x_1+y)\psi_-(y)e^{\kappa x_1}
,\;\;x_1 \geq 0\cr
D\phi_+(y)\phi_-(x_1+y) \equiv D\psi_+(y)\psi_-(x_1+y)e^{-\kappa x_1},\;\;x_1\leq 0.}
\end{equation}

Since they are periodic, the functions $\psi_\pm$ can be expressed as Fourier series:
\begin{eqnarray}
\psi_+(x_1) &=& \sum_{m=-\infty}^\infty\,a_m(\kappa)e^{-2\pi im x_1/h},\\
\psi_-(x_1) &=& \sum_{m=-\infty}^\infty\,a_m(-\kappa)e^{-2\pi im x_1/h},
\end{eqnarray}
where
\begin{eqnarray}
a_m(\kappa) &=& \frac{1}{h}\int_0^h\,\psi_+(x_1)e^{2\pi im x_1/h}\,dx_1
\equiv \frac{1}{h}\int_0^h\,\phi_+(x_1)e^{(2\pi im/h -\kappa)x_1}\,dx_1,\\
a_m(-\kappa) &=& \frac{1}{h}\int_0^h\,\psi_-(x_1)e^{2\pi im x_1/h}\,dx_1
\equiv \frac{1}{h}\int_0^h\,\phi_-(x_1)e^{(2\pi im/h +\kappa)x_1}\,dx_1.
\end{eqnarray}

\section{Effective properties}

Green's function for the effective medium is the ensemble average:
\begin{eqnarray}
\langle G\rangle(x_1) &=& \frac{1}{h}\int_0^h\,G(x_1,y)\,dy\nonumber\\
&=&\cases{D e^{\kappa x_1}\sum_m\,a_m(\kappa)a_{-m}(-\kappa)e^{-2\pi i m x_1/h},\;\;x_1 \geq 0\,\cr
D e^{-\kappa x_1}\sum_m\,a_m(\kappa)a_{-m}(-\kappa)e^{2\pi i m x_1/h},\;\;x_1 \leq 0}\nonumber\\
&=&De^{\kappa \vert x_1\vert}\sum_m\,a_m(\kappa)a_{-m}(-\kappa)e^{-2\pi i m \vert x_1\vert/h}.
\end{eqnarray}

Stresses $\sigma_{i3}$ and momentum $p$ can be treated similarly. First, the exact stress component
$\sigma_{13}$ is
\begin{equation}
\sigma_{13}(x_1,y) = \cases{D\mu^0_{13}(x_1+y)\phi_+^\prime(x_1+y)\phi_-(y),\;\;x_1 > 0,\cr
    D\mu^0(x_1+y)\phi_+(y)\phi_-^\prime(x_1+y),\;\;x_1 < 0.}
\end{equation}
Now with
\begin{equation}
\mu_{13}^0(x_1)\phi_+^\prime(x_1)e^{-\kappa x_1} = \sum_{m=-\infty}^\infty\,b_m(\kappa)e^{-2\pi imx_1/h}
\end{equation}
and
\begin{equation}
\mu^0_{13}(x_1)\phi_-^\prime(x_1)e^{\kappa x_1} = \sum_{m=-\infty}^\infty\,b_m(-\kappa)e^{-2\pi imx_1/h},
\end{equation}
the ensemble averaged stress component can be expressed
\begin{equation}
\langle\sigma_{13}\rangle(x_1) = \cases{D e^{\kappa x_1}\sum_m\,b_m(\kappa)a_{-m}(-\kappa)
e^{-2\pi im x_1/h},\;\;x_1 > 0,\cr
D e^{-\kappa x_1}\sum_m\,a_{m}(\kappa)b_{-m}(-\kappa)e^{2\pi im x_1/h},\;\;x_1 < 0.}
\end{equation}
Similarly,
\begin{equation}
\langle\sigma_{23}\rangle(x_1) = \cases{Dk e^{\kappa x_1}\sum_m\,c_m(\kappa)a_{-m}(-\kappa)
e^{-2\pi im x_1/h},\;\;x_1 > 0,\cr
Dk e^{-\kappa x_1}\sum_m\,a_{m}(\kappa)c_{-m}(-\kappa)e^{2\pi im x_1/h},\;\;x_1 < 0}
\end{equation}
and
\begin{equation}
\langle p\rangle(x_1) = \cases{s D e^{\kappa x_1}\sum_m\,d_m(\kappa)a_{-m}(-\kappa)
e^{-2\pi im x_1/h},\;\;x_1 > 0,\cr
s D e^{-\kappa x_1}\sum_m\,a_{m}(\kappa)d_{-m}(-\kappa)e^{2\pi im x_1/h},\;\;x_1 < 0,}
\end{equation}
where
\begin{equation}
\mu_{23}^0(x_1)\phi_+(x_1)e^{-\kappa x_1} = \sum_{m=-\infty}^\infty\,c_m(\kappa)e^{-2\pi imx_1/h},
\end{equation}
\begin{equation}
\mu^0_{23}(x_1)\phi_-(x_1)e^{\kappa x_1} = \sum_{m=-\infty}^\infty\,c_m(-\kappa)e^{-2\pi imx_1/h},
\end{equation}
\begin{equation}
\rho^0(x_1)\phi_+(x_1)e^{-\kappa x_1} = \sum_{m=-\infty}^\infty\,d_m(\kappa)e^{-2\pi imx_1/h}
\end{equation}
and
\begin{equation}
\rho^0(x_1)\phi_-(x_1)e^{\kappa x_1} = \sum_{m=-\infty}^\infty\,d_m(-\kappa)e^{-2\pi imx_1/h}.
\end{equation}

\subsection*{Fourier transforms}

Now take
\begin{equation}
s = \epsilon - i\omega \;\hbox{ and }\;k = -i\xi_2,
\end{equation}
where $\epsilon$ is small and positive, and Fourier transform the preceding ensemble averages
with respect to $x_1$. This gives
\begin{equation}
\widetilde{\langle G\rangle}(\xi_1,\xi_2,\omega)= -2D\sum_{m=-\infty}^\infty\,
\frac{a_m(\kappa)a_{-m}(-\kappa)(\kappa-2\pi i m/h)}{(\kappa - 2\pi i m/h)^2 + \xi_1^2},
\end{equation}
\begin{equation}
\widetilde{\langle\Sigma_{13}\rangle}(\xi_1,\xi_2,\omega) = -D\sum_m\,\left\{
\frac{a_m(\kappa)b_{-m}(-\kappa)}{\kappa - 2\pi i m/h -i\xi_1}
+\frac{b_m(\kappa)a_{-m}(-\kappa)}{\kappa - 2\pi i m/h - i\xi_1}\right\},
\end{equation}
\begin{equation}
\widetilde{\langle\Sigma_{23}\rangle}(\xi_1,\xi_2,\omega) = i\xi_2 D\sum_m\,\left\{
\frac{a_m(\kappa)c_{-m}(-\kappa)}{\kappa - 2\pi i m/h -i\xi_1}
+\frac{c_m(\kappa)a_{-m}(-\kappa)}{\kappa - 2\pi i m/h +i\xi_1}\right\},
\end{equation}
\begin{equation}
\widetilde{\langle P\rangle}(\xi_1,\xi_2,\omega) = i\omega D\sum_m\,\left\{
\frac{a_m(\kappa)d_{-m}(-\kappa)}{\kappa - 2\pi i m/h -i\xi_1}
+\frac{d_m(\kappa)a_{-m}(-\kappa)}{\kappa - 2\pi i m/h +i\xi_1}\right\}.
\end{equation}
Upper-case $\Sigma$ and $P$ are employed, to identify these quantities as components of stress
and momentum density associated with the Green's function.
Note that, in these expressions, $\kappa$ is a function of $\xi_2$ and $\omega$.

\subsection*{Effective constitutive relations}

Following \cite{JRW09}, equation (29) can be written
\begin{equation}
\widetilde{\langle\Sigma_{13}\rangle} = \tilde C^{eff}_{1313}(-i\xi_1\widetilde{\langle G\rangle})
+ \tilde S^{eff}_{133}(-i\omega\widetilde{\langle G\rangle}),
\end{equation}
where
\begin{equation}
\tilde C^{eff}_{1313} = D\sum_m\,\frac{a_m(\kappa)b_{-m}(-\kappa)-a_{-m}(-\kappa)b_m(\kappa)}
{(\kappa-2\pi i m/h)^2 +\xi_1^2}\widetilde{\langle G\rangle}^{-1},
\end{equation}
\begin{equation}
\tilde S^{eff}_{133} = -D\sum_m\,\frac{[a_m(\kappa)b_{-m}(-\kappa)+a_{-m}(-\kappa)b_m(\kappa)]
(\kappa - 2\pi i m/h)}{(\kappa-2\pi i m/h)^2 +\xi_1^2}\widetilde{\langle G\rangle}^{-1}.
\end{equation}

There is a correspondingly ``obvious'' split for equation (31), whereas a similar split for (30)
gives a less ``intuitive'' result. There is, however, the general formula for effective properties
\begin{equation}
{\cal L}^{eff} = \langle{\cal L}\rangle - \langle{\cal L}{\cal E}({\cal E}G^\dag)^\dag{\cal L}\rangle
+\langle{\cal L}{\cal E}G\rangle\langle G\rangle^{-1}\langle({\cal E}G^\dag)^\dag{\cal L}\rangle
\end{equation}
derived in \cite{JRW12b}, which applies also in the presence of any non-random inelastic
strain.\footnote{The formula as given in \cite{JRW12b} was more general because it
allowed also for a ``weighted
average'', not considered at present.} In the present context, ${\cal L}$ is the
$3\times 3$ array of parameters\footnote{${\cal L}$ is diagonal but
${\cal L}^{eff}$ is not.} that relates
$${\bf s} = \left(\matrix{\sigma_{13}\cr
                          \sigma_{23}\cr
                          p\cr}\right) 
                          \;\hbox{ to }\;
  {\cal E}u_3 = \left(\matrix{u_{3,1}\cr
                              u_{3,2}\cr
                              \dot u_3\cr}\right). $$
                              
Implementing (35) in the Fourier transform domain gives
\begin{equation}
\tilde{\cal L}^{eff} = \left(\matrix{\tilde C_{1313}^{eff}&C_{1323}^{eff} &
\tilde S_{133}^{eff}\cr
\tilde C_{2313}^{eff} & C_{2323}^{eff} & \tilde S_{233}^{eff}\cr
\tilde S_{313}^{eff} & \tilde S_{123}^{eff} & \tilde \rho_{33}^{eff}\cr}\right).
\end{equation}

The term $\langle{\cal L}\rangle -\langle{\cal L}{\cal E}({\cal E} G^\dag)^\dag{\cal L}\rangle$ is
significant in the presence of inelastic deformation but makes zero contribution to the stress
when strain and velocity are derived from a displacement. However, for the record, the required
formula is
\begin{eqnarray}
\widetilde{\langle{\cal L}\rangle}
 -\widetilde{\langle{\cal L}{\cal E}({\cal E} G^\dag)^\dag{\cal L}\rangle} &=&
\left(\matrix{0 & 0 & 0\cr
                                           0&\widetilde{\langle{C_{2323}}\rangle} & 0\cr                                           0&0&\widetilde{\langle\rho\rangle}\cr}\right)\nonumber\\
                                           &&\hbox{\hskip -1.75in}
+D\sum_m\,\frac{1}{\kappa -2\pi i m/h -i\xi_1}
\left(\matrix{b_{-m}(-\kappa)\cr
       -i\xi_2c_{-m}(-\kappa)\cr
       -i\omega d_{-m}(-\kappa)\cr}\right)
       \left(\matrix{b_m(\kappa)& i\xi_2 c_m(\kappa)&-i\omega d_m(\kappa)\cr}\right)\nonumber\\
&&\hbox{\hskip -1.75in}
+D\sum_m\,\frac{1}{\kappa -2\pi i m/h +i\xi_1}\left(\matrix{b_m(\kappa)\cr
       -i\xi_2c_m(\kappa)\cr
       -i\omega d_m(\kappa)\cr}\right)
       \left(\matrix{b_{-m}(-\kappa)& i\xi_2 c_{-m}(-\kappa)&-i\omega d_{-m}(-\kappa)\cr}\right).
\nonumber\\
\end{eqnarray}
The reason for the absence of a term $\widetilde{\langle C_{1313}\rangle}$ in the $(1,1)$
place in the first matrix
is that it is cancelled by a delta-function contribution from ${\cal E}({\cal E}G^\dag)^\dag$ in that position.

The remaining term in (35) can be built up by noting that
\begin{equation}
\widetilde{\langle{\cal L}{\cal E}G\rangle}(\xi_1,\xi_2,\omega)
=\left(\matrix{\widetilde{\langle\Sigma_{13}\rangle}(\xi_1,\xi_2,\omega)\cr
               \widetilde{\langle\Sigma_{23}\rangle}(\xi_1,\xi_2,\omega)\cr
               \widetilde{\langle P\rangle}(\xi_1,\xi_2,\omega)\cr}\right)
\end{equation}
where $\widetilde{\langle\Sigma_{13}\rangle}$ etc. are given respectively by (29), (30) and (31),
and that
\begin{equation}
\widetilde{\langle({\cal E}G^\dag)^\dag{\cal L}\rangle}(\xi_1,\xi_2,\omega)
= \widetilde{\langle{\cal L}{\cal E}G\rangle}^\dag(\xi_1,\xi_2,\omega)
=\{\widetilde{\langle{\cal L}{\cal E}G\rangle}(-\xi_1,-\xi_2,\omega)\}^T,
\end{equation}
while $\widetilde{\langle G\rangle}$ is given by(28). Note that $\kappa$ depends on
$\xi_2$ and $\omega$ and is an even function of $\xi_2$.

\subsubsection*{Some elementary observations}

Note first that equation (16) represents $\langle G\rangle$ as a superposition of plane waves,
with space and time-dependence $e^{st+(\kappa-2\pi i m/h)x_1 + kx_2}$ when $x_1 >0$
and a corresponding dependence with $\kappa$ replaced by $-\kappa$ when $x_1< 0$.
Each of these of course corresponds to a single ``Floquet--Bloch'' mode (depending only on the
sign before $\kappa$). Another view of the same result is obtained from considering the Fourier
transform of $\langle G\rangle$. Poles of the transform correspond to plane waves, and
the poles are at points $(\xi_1,\xi_2,\omega)$ at which 
\begin{equation}
i\xi_1 \pm (\kappa(-i\xi_2,-i\omega)-2\pi i m/h) = 0.
\end{equation}
Furthermore, the expression for $\widetilde{\langle{\cal L}\rangle}$ as $(\xi_1,\xi_2,\omega)$
approaches one of the poles involves only the contributions from that pole. Thus, for instance,
\begin{equation}
\widetilde{\langle\Sigma_{13}\rangle} \sim 
-D\frac{b_m(\kappa)a_{-m}(-\kappa)}{\kappa - 2\pi i m/h +i\xi_1}
\end{equation}
as $i\xi_1 + (\kappa(-i\xi_2,-i\omega)-2\pi i m/h) \to 0$. The expression for
$\widetilde{\langle{\cal L}\rangle}$ simplifies correspondingly, and clearly depends
on $m$.

Next, a comment on Green's function in physical space. In equation (16), the dependence on $x_1$
is explicit but (even assuming that harmonic time-dependence is required) it is necessary to
invert the Laplace transform with respect to $k$. Alternatively, it is perhaps simpler
to start from the Fourier transform (28) and invert first with respect to $\xi_2$. Poles
in (say) the upper half of the complex $\xi_2$-plane are certain to appear wherever
\begin{equation}
\kappa(-i\xi_2,-i(\omega+ 0i)) = 2\pi i m/h -i\xi_1.
\end{equation}
I am inclined to expect that there will be one solution for each $m$, but have not properly
investigated; nor do I know if there are other (branch-cut) singularities. I do intend to
investigate this further, in order to calculate the influence of a line of body-force,
proportional to $e^{i(\omega t + \xi_1 x_1)}\delta(x_2)$. This will be of interest in relation
to considering transmission and reflection from a medium occupying a half-space $x_2 >0$.

\section{A note on implementation}

Consider an $n$-phase laminate, in which material of type $j$ has elastic constants
$C^j_{13},\; C^j_{23}$ and density $\rho^j$, and
occupies the region $z_{j-1} < x_1 < z_j$, where $z_0=0$ and $z_j = z_{j-1} + h_j$.
This structure is repeated periodically, with
period $h = \sum_{j=1}^n\,h_j$. To obtain the basic Floquet waves $\phi_\pm$, $A_j$ is defined to be
\begin{equation}
A_j = \left(\matrix{\sigma_{13}(z_{j-1})\cr
                    u_3(z_{j-1})\cr}\right).
\end{equation}
Then, for $z_{j-1} < x_1 < z_j$,
\begin{equation}
\left(\matrix{\sigma_{13}(x_1)\cr
                u_3(x_1)\cr}\right) = M_j(x_1-z_{j-1})A_j,
\end{equation}
where the propagator matrix $M_j(x_1-z_{j-1})$ is given as
\begin{equation}
M_j(x_1-z_{j-1}) = \left(\matrix{\cosh(\kappa_j(x_1-z_{j-1}))& \mu^j_{13}\kappa_j\sinh(\kappa_j(x_1-z_{j-1}))\cr
(\mu^j_{13}\kappa_j)^{-1}\sinh(\kappa_j(x_1-z_{j-1})) & \cosh(\kappa_j(x_1-z_{j-1}))\cr}\right),
\end{equation}
with
\begin{equation}
\kappa_j = \sqrt{\frac{\rho^j s^2-\mu^j_{23}k^2}{\mu^j_{13}}}.
\end{equation}
In particular,
\begin{equation}
A_{j+1} = M_j(h_j)A_j,
\end{equation}
and the Floquet condition that fixes $\kappa$ is defined so that
\begin{equation}
M A_1 = e^{\kappa h}A_1,
\end{equation}
where
\begin{equation}
M = M_1(h_1)M_2(h_2)\cdots M_n(h_n).
\end{equation}
It may be noted, in passing, that $\det(M) =1$, because each $M_j(h_j)$ has determinant $1$,
and hence the Floquet condition can be written as
\begin{equation}
\cosh(\kappa h) = \half{\rm trace}(M).
\end{equation}
The stress component $\sigma_{23}$ will also be needed. When $z_{j-1} < x_1 < z_j$,
\begin{equation}
\sigma_{23}(x_1) = \mu^j_{23}k\left[\matrix{
(\mu^j_{13}\kappa_j)^{-1}\sinh(\kappa_j(x_1-z_{j-1})), & \cosh(\kappa_j(x_1-z_{j-1}))\cr}\right]A_j.
\end{equation}
Now all that remains is the tedium of actually doing the calculations...


\begin{thebibliography}{22}
\bibitem{JRW09} J.R. Willis. Exact effective relations for dynamics of a laminated body. Mechanics of Materials {\bf 41} (2009), 385-393.
\bibitem{JRW12} J.R. Willis. A comparison of two formulations for effective relations for waves in a composite. Mechanics of Materials {\bf 47} (2012), 51-60.
\bibitem{JRW12b} J.R. Willis. The construction of effective relations for waves in a composite.
Comptes Rendus M\'ecanique {\bf 340} (2012), 181-192. 
\end{thebibliography}
\end{document}